\documentclass[showpacs,amsmath,amssymb,twocolumn,prl,superscriptaddress]{revtex4-1}
\usepackage{amssymb}
\usepackage[dvips]{graphicx}
\usepackage{enumerate}
\usepackage{epsfig}
\usepackage{subfigure}
\usepackage{xcolor}
\usepackage[T1]{fontenc}
\usepackage{fullpage}
\usepackage{amsthm,amsfonts,amssymb,amscd,mathrsfs,xspace,framed}
\usepackage{amsmath}
\usepackage{color}
\usepackage{setspace}
\usepackage{url}
\usepackage{wrapfig}
\usepackage{enumitem}
\begin{document}

\title{Efficient net-gain integrated optical parametric amplifier in the quantum regime}

\author{Yung-Cheng Kao}
\affiliation{Chandra Department of Electrical and Computer Engineering, The University of Texas at Austin, Austin, Texas 78758, USA}
\author{Jiaqi Huang}
\affiliation{Chandra Department of Electrical and Computer Engineering, The University of Texas at Austin, Austin, Texas 78758, USA}
\author{Ian Briggs}
\affiliation{Chandra Department of Electrical and Computer Engineering, The University of Texas at Austin, Austin, Texas 78758, USA}
\author{Pao-Kang Chen}
\affiliation{Chandra Department of Electrical and Computer Engineering, The University of Texas at Austin, Austin, Texas 78758, USA}
\author{Linran Fan}
\email{linran.fan@utexas.edu}
\affiliation{Chandra Department of Electrical and Computer Engineering, The University of Texas at Austin, Austin, Texas 78758, USA}

\maketitle

\textbf{
Optical parametric amplifiers (OPAs) are promising to overcome the wavelength coverage and noise limitations in conventional optical amplifiers based on rare-earth doping and semiconductor gain.
However, the high power requirement remains a major obstacle to the widespread use of OPAs. 
Integrated OPAs can in principle improve the pump efficiency with tight mode confinement; however, challenges associated with propagation loss, limited nonlinearity, and susceptibility to nanoscale fabrication imperfections prevent them from competing with conventional bulk and fiber-based OPAs.
Here, we demonstrate a highly efficient integrated OPAs with continuous-wave net gain. The pump efficiency is improved by over one order of magnitude. Phase-sensitive gain of 23.5~dB is demonstrated, significantly exceeding previous integrated OPAs, using only 110~mW pump power and no cavity enhancement. 
This is achieved with parametric down-conversion in thin-film lithium niobate waveguides using the adapted poling technique to maintain the coherence of nonlinear interactions.
Moreover, the high parametric gain exceeds fibre–chip–fibre losses, leading to appreciable net gain up to 10~dB.
The 3-dB bandwidth is approximately 120 nm, covering telecommunication S-, C-, and L-bands.
Quantum-limited noise performance is confirmed through the measurement of output field fluctuation below the classical limit.
We further demonstrate that signal-to-noise ratio in noisy optical communications can be increased by leveraging this efficient integrated OPA.
Our work marks a significant step towards ideal optical amplifiers with strong amplification, high efficiency, quantum-limited noise, large bandwidth, and continuous-wave operation, unlocking new possibilities for next‑generation photonic information processing systems.
}

\begin{figure*}
\centering
\includegraphics[width=0.8\textwidth]{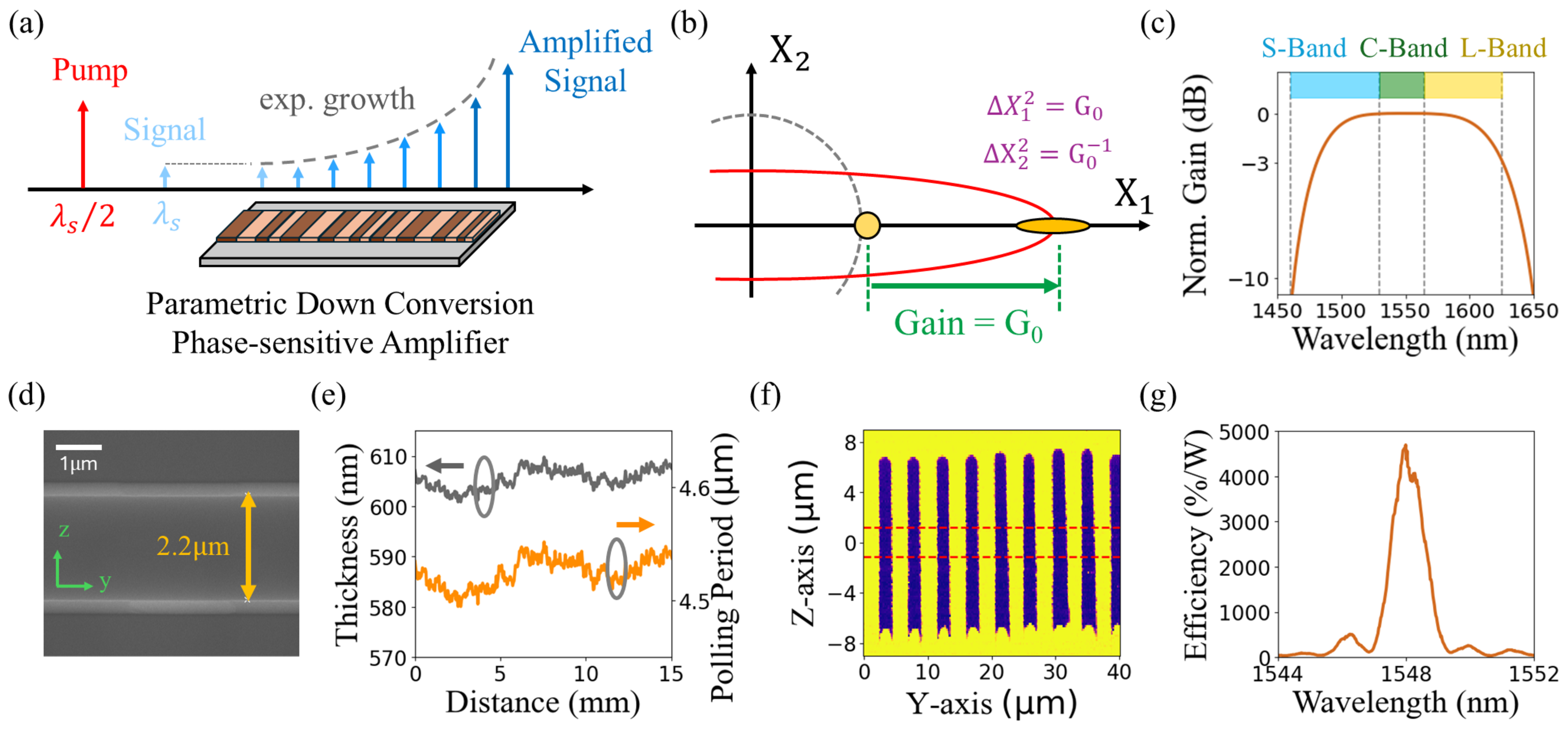}
\caption{
\textbf{Integrated OPA with TFLN waveguides using adapted poling.}
\textbf{a}, Schematic of OPAs based on parametric down-conversion in single-pass waveguides. In the degenerate phase-sensitive case, the pump wavelength is half of the signal wavelength. The signal is exponentially amplified along the waveguide.
\textbf{b}, Phase dependence of OPA gain. The gain is maximized when the phase difference between signal and pump is zero (in-phase), and minimized when the phase difference is $\pi/2$ (out-of-phase).
\textbf{c}, Simulated gain spectrum of a 14-mm long integrated OPA. The simulated 3-dB bandwidth is around 140~nm, covering the S-, C- and L-band for telecommunication. 
\textbf{d}, Scanning electron microscopy (SEM) image of fabricated TFLN waveguides with 2.2~$\mu$m width. Green arrows indicate the crystal direction of lithium niobate.
\textbf{e}, Measured film thickness (grey) and designed poling period (orange) along TFLN waveguides. The poling period is adjusted based on the measure film thickness to ensure perfect phase-matching along the waveguide.
\textbf{f}, Piezoelectric force microscopy (PFM) image of the lithium niobate poling regime.
\textbf{g}, Measured second-harmonic generation spectrum of TFLN waveguides with adapted poling. The spectrum shows a nearly theoretical sinc-squared shape with peak nonlinear efficiency 4700 $\pm$ 500~\%/W}
\label{fig1}
\end{figure*}

Optical amplification is of pivotal importance in modern communications and information processing~\cite{Winzer2018,Zimmerman2004,Willner2014}. 
Optical amplifiers are generally assessed in terms of gain, noise performance, and bandwidth.
Currently, the dominant optical amplifier technologies rely on electronic transitions in rare-earth dopants~\cite{Desurvire1987} and semiconductors~\cite{Michael2007}. However, their bandwidth is constrained by the underlying electronic energy levels, while their noise performance is fundamentally limited by spontaneous emission~\cite{Caves1982}. 
Optical parametric amplifiers (OPAs) provide a promising route to realize ultimate optical amplification performance with high gain, large bandwidth, tunable center wavelength, and low noise at the same time.
High-performance OPAs have been widely demonstrated using Kerr nonlinearity in optical fibers~\cite{Marhic1996,Hansryd2001} and Pockels nonlinearity in bulk micromachined lithium niobate~\cite{Umeki2011,Asobe2012,Kishimoto2016,Sua2018,Kazama2021}, enabling applications in optical communications~\cite{Tong2011,Olsson2018}, ultrafast signal processing~\cite{Slavik2010,Marhic2014,Willner2014,Takanashi2020}, and quantum technology~\cite{aasi2013enhanced, Hudelist2014}.

Despite their superior performance metrics and pioneering demonstrations, the widespread deployment of OPAs has been severely constrained by high pump power requirements. Photonic integrated circuits offer a promising path toward device miniaturization and reduced power consumption~\cite{Ferrera2008,Miller2017}. However, realizing high-performance OPAs on chip remains challenging. Appreciable gain still requires high pump power due to unique challenges in photonic integrated circuits such as high propagation loss, limitation of material nonlinearity, and poor tolerance of nanoscale variations. Thus, integrated OPAs have not shown obvious performance advantages compared with conventional OPAs based on fiber or bulk lithium niobate~\cite{Hansryd2001,Torounidis2006,Asobe2012, Umeki2011,Kishimoto2016,Sua2018,Kazama2021,Ye2021,Riemensberger2022,Kuznetsov2025,Chen2025,Zhao2025,Dean2026}.

Here, we address this challenge and demonstrate an integrated OPA that delivers substantial net gain up to 10~dB together with more than an order‑of‑magnitude improvement in pump efficiency. This is achieved without any cavity enhancement, ensuring continuous broadband coverage. Our approach leverages low‑loss thin‑film lithium niobate waveguides, which provide strong $\chi^{(2)}$ nonlinearity and tight optical confinement (Fig.~\ref{fig1}a). In particular, we employ the recently developed adapted poling technique to mitigate nanoscale inhomogeneities that have long limited the nonlinear efficiency of thin‑film lithium niobate waveguides~\cite{Chen2023}.

\begin{figure*}[ht]
\centering
\includegraphics[width=0.98\textwidth]{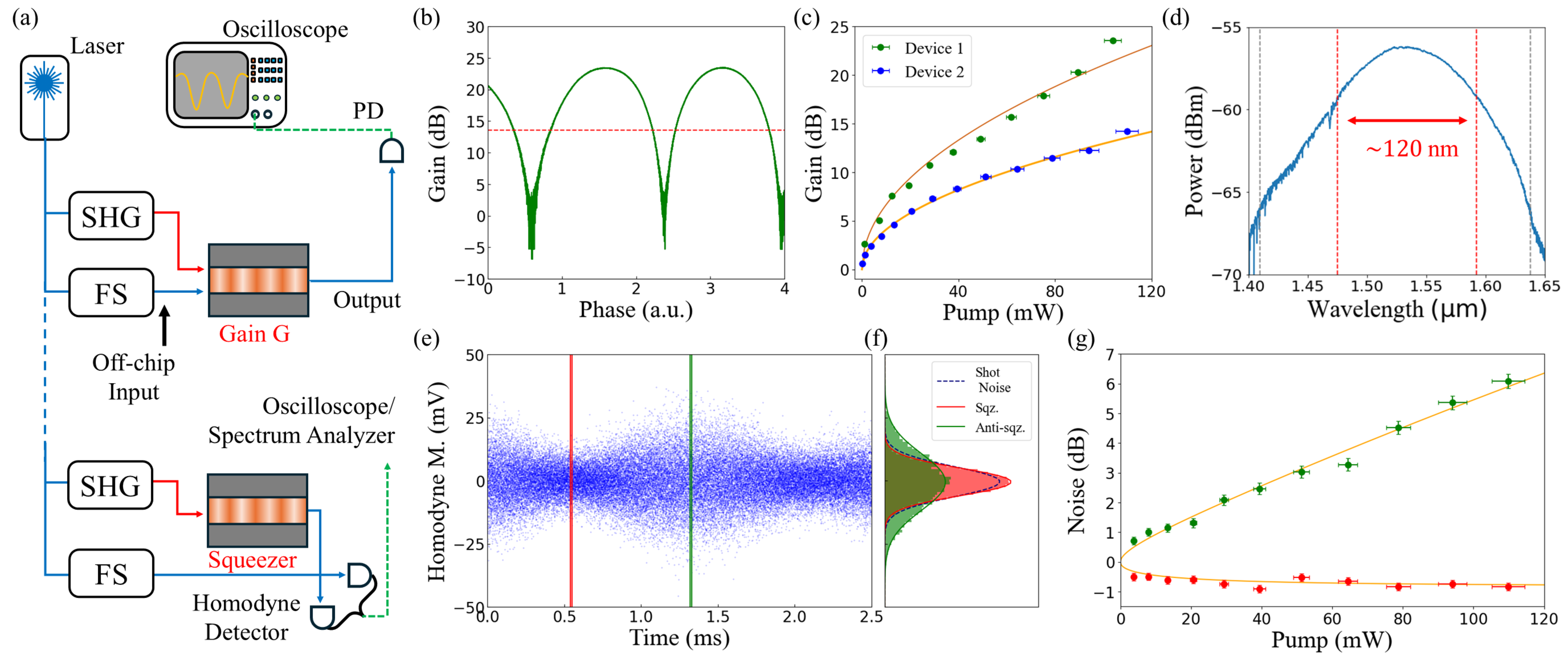}
\caption{
\textbf{Performance characterization of the integrated OPA.}
\textbf{a}, Experimental setup to characterize the integrated OPA. The upper and lower panels show the direct power measurement and homodyne detection respectively. SHG: commercial second-harmonig generation module; FS: fiber strecher, PD: photodetector. 
\textbf{b}, Measured OPA gain with phase scanning with 110~mW pump power. The output signal power without pump light used as the reference (0~dB). The red dashed line (13.5~dB) indicates the fiber-chip-fiber transmission loss. 
\textbf{c}, Measured integrated OPA gain as a function of pump power.
\textbf{d}, Measured spontaneous emission spectrum of the integrated OPA with pump power 110~mW, showing 3-dB bandwidth over 120~nm.
\textbf{e}, Time-domain homodyne measurement of output field quadrature under phase scan. The red (green) lines mark the squeezing (anti-squeezing) conditions. The sampling rate is 2~GS/s. 
\textbf{f}, Statistical distributions of time-domain homodyne measurement over 10~$\mu$s (20,000 samples) for shot noise with the pump blocked, squeezing (red line in \textbf{e}), and anti-squeezing (green line in \textbf{e}). The distributions are fitted with Gaussian functions to extract the variances (See methods).
\textbf{g}, Measured squeezing and anti-squeezing level as a function of pump power. Uncertainties in pump power in \textbf{c} and \textbf{g} is determined by the variance of measured coupling efficiencies in different devices on the chip. Uncertainties in gain value in \textbf{c} and noise level in \textbf{g} are evaluated by the standard deviation of the fitted parameters. }
\label{fig2}
\end{figure*}

The phase-sensitive gain of OPAs based on $\chi^{(2)}$ nonlinearity is expressed as~\cite{Caves1982}
\begin{equation}
\begin{aligned}
& G = \cos^2\theta ~G_0+\sin^2\theta ~G_0^{-1}
\label{Eq:PSA_gain}
\end{aligned}
\end{equation}
where $\theta$ is the phase between the pump and signal, $G_0=\exp(2\sqrt{\eta P_\mathrm{p}})$ is the gain at zero phase, $\eta$ is the nonlinear efficiency, and $P_\mathrm{p}$ is the pump power. 
The phase-sensitive nature of OPA determines the maximum gain can only be achieved for the in-phase quadrature of the signal, and the out-of-phase quadrature is de-amplified (Fig.~\ref{fig1}b).
The TFLN waveguide is designed as 14-mm long and 2.2-$\mu$m wide, with the phase matching condition engineered for the amplification near 1550~nm with the pump near 775~nm. 
Because of the small dispersion, we expect the integrated OPA can cover the S-, C-, and L-band for telecommunication (Fig.~\ref{fig1}c, see Methods).
Because the nonlinear efficiency scales quadratically with device length ($\eta \propto L^2$), the resulting gain grows exponentially as the length increases~(Fig.~\ref{fig1}a). However, nanoscale thickness variation intrinsic to thin-film lithium niobate limit the coherent interaction length of $\chi^{(2)}$ nonlinear processes to a few millimeters~\cite{Wang2018a,Rao2019,Zhao2020,Chen2022,Chang2016,Boes2019,Zhang2022}. With conventional periodic poling, $\chi^{(2)}$ processes acquire random phases along TFLN waveguides due to local geometric variations, which prevents constructive signal buildup and reduces nonlinear efficiency. Therefore, we adjust poling periods depending on local thickness variations to recover the ideal phase matching condition~(Fig.~\ref{fig1}d-f).
The recovery of the ideal phase matching condition is verified by the second-harmonic generation spectrum that matches the theoretical sinc-squared function with nonlinear efficiency of 4700 $\pm$ 500~\%/W (Fig.~\ref{fig1}f, Supplementary Section~1).

Two methods are employed to characterize the performance of the integrated OPA including the direct detection of power and the homodyne detection of field amplitude (Fig.~\ref{fig2}a).
In the direct power measurement, the signal light near 1550 nm and the pump light near 775 nm are coupled into the integrated OPA together. The phase between the signal and pump light is scanned by the fiber stretcher, therefore the output power of the signal light oscillates between amplification and de-amplification~(Fig.~\ref{fig2}b). The output power of the signal light with the pump turned off is used as the reference to calculate the on-chip gain. With the pump power at 110~mW, we measure phase-sensitive gain of 23.5~dB, significantly exceeding previous integrated OPAs~(Fig.~\ref{fig2}b and c, see Methods)~\cite{Ye2021,Riemensberger2022,Kuznetsov2025,Chen2025,Zhao2025,Dean2026}. This corresponds to a pump efficiency up to 15.3~dB/(W$\cdot$mm). This is one order of magnitude higher than previous integrated and bulk OPAs (Supplementary Section 3).
We further measure the input power of the signal light (red dashed line in Fig.~\ref{fig2}b), which is 13.5~dB higher than the reference output power. This shows that our integrated OPA can achieve over 10~dB net gain. The measured gain dependence on the pump power follows Eq.~\ref{Eq:PSA_gain}. We further measure a second device with a lower nonlinear efficiency (Supplementary Section~1). Gain over 15~dB is achieved , also achieving net gain (Fig.~\ref{fig2}c). 
The 3-dB gain bandwidth is around 120~nm from 1470~nm to 1590~nm (Fig.~\ref{fig2}d), significantly exceeding the bandwidth of conventional erbium-doped fiber amplifiers.


\begin{figure*}[ht]
\centering
\includegraphics[width=0.98\textwidth]{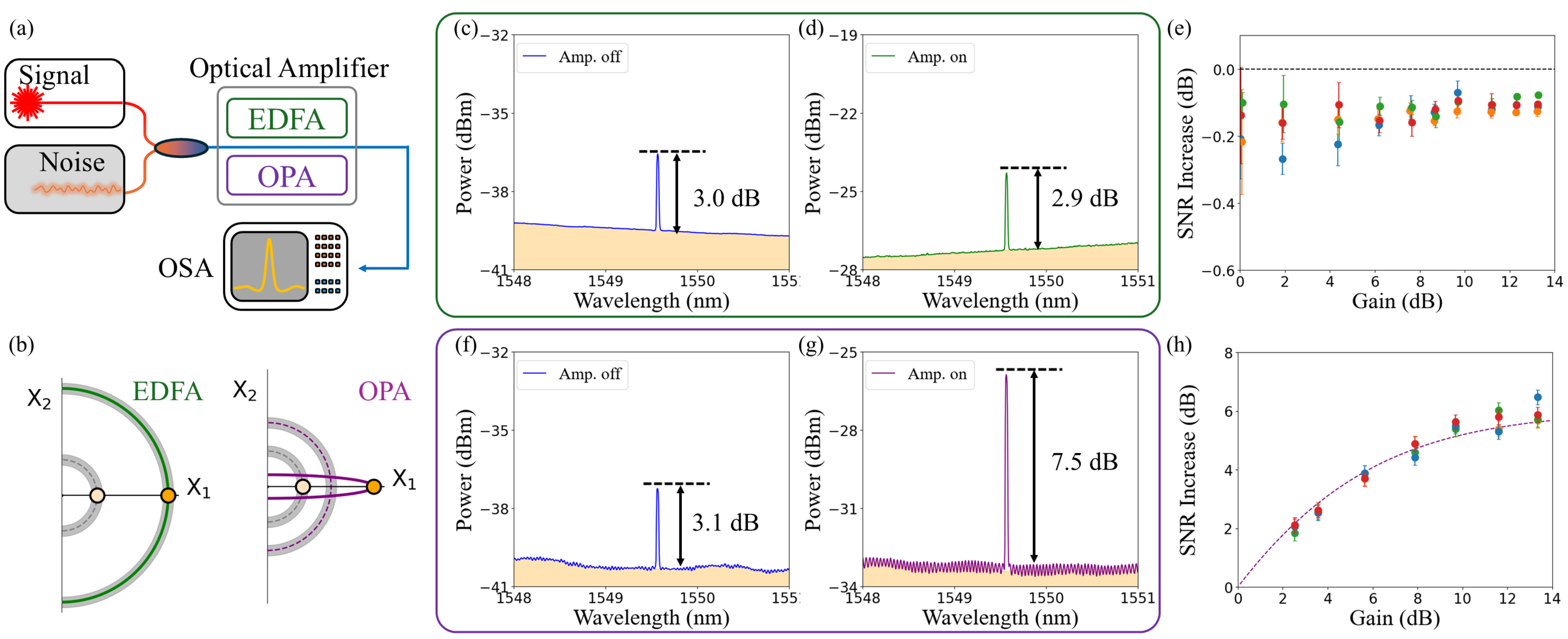}
\caption{
\textbf{Comparison between EDFA and integrated OPA for signal-to-noise ratio improvement}
\textbf{a}, Experimental schematic. A single-frequency laser at 1549.6 nm is mixed with broadband noise generated from unseeded EDFA. After amplification with EDFA or OPA, the output light is detected by an optical spectrum analyzer (OSA).
\textbf{b}, Phase diagram for EDFA and OPA amplification. The signals before and after amplification are labeled by orange circles respectively, and grey areas show the noise. EDFA amplify the signal and noise uniformly for all phases. OPA selectively amplifies in-phase signal and noise and de-amplifies out-of-phase noise.
\textbf{c, d}, Optical spectrum before and after 13~dB EDFA amplification.
\textbf{e}, SNR change after EDFA amplification with respect to gain. The SNR change has near-zero dependence on the gain, and shows an average of -0.15~dB. 
\textbf{f, g}, Optical spectrum before and after 13~dB OPA amplification.
\textbf{h}, SNR change after OPA amplification with respect to gain. The SNR change increases with the OPA gain, and approaches 6~dB. Different optical SNR values before amplification are tested in \textbf{e} and \textbf{h} including 0~dB (blue), 5~dB (orange), 8~dB (green), and 12~dB (red).
Uncertainties in \textbf{e} and \textbf{h} are obtained by calculating the variance of repeated measurements.
}
\label{fig3}
\end{figure*}

To test the integrated OPA in the quantum regime, we replace the direct power measurement with homodyne detection (Fig.~\ref{fig2}a). The phase of the local-oscillator is scanned with the fiber stretcher to measure the output signal light along different quadratures. An oscilloscope is used to sample the homodyne detection output and extract the quadrature variance. 
We observe clear periodic increase and decrease of the quadrature variance, corresponding to the squeezing and anti-squeezing of the classical shot noise respectively (Fig.~\ref{fig2}e). The shot noise is calibrated by blocking the pump light, and the squeezing and anti-squeezing level can be obtained through the ratio between the field variance of the output signal and the shot noise reference (Fig.~\ref{fig2}f) (See Methods).
As shown in Fig.~\ref{fig2}g, both the squeezing and anti-squeezing levels increase with the pump power, and a maximum squeezing level of 0.8~dB below the classical shot-noise limit is measured with a pump power of 110~mW.


To further demonstrate the capability of the integrated OPA, we compare its performance with EDFA in noisy optical systems. 
The optical input signal is mixed with incoherent noise, and amplified with the integrated OPA or EDFA. The output signal is analyzed with the optical spectrum analyzer (Fig.~\ref{fig3}a). The EDFA amplifies both the signal and noise uniformly. In contrast, the integrated OPA selectively amplifies light that is in-phase with the pump and along the phase-matching polarization (Fig.~\ref{fig3}b). 
To compare the EDFA and integrated OPA performance, we measure the signal-noise ratio (SNR) before and after amplification. The input signal and noise powers are set equal at the input in both cases (Fig.~\ref{fig3}c and f). Because of the uniform amplification of signal and noise, the SNR after EDFA amplification does not change and has near-zero dependence on the both the EDFA gain (Fig.~\ref{fig3}d and e). On the other hand, significant SNR improvement is observed after OPA amplification (Fig.~\ref{fig3}g). Moreover, the SNR improvement approaches 6~dB as the OPA gain increase (Fig.~\ref{fig3}h), following the relation (Supplementary Section~2)
\begin{equation}
\begin{aligned}
    \mathrm{\frac{SNR_{out}}{SNR_{in}}} = \frac{4G_0}{2+G_0+G_0 ^{-1}}\rightarrow 6~\mathrm{dB} 
\end{aligned}
\label{eq:snr_change}
\end{equation}


\begin{figure*}[ht]
\centering
\includegraphics[width=0.98\textwidth]{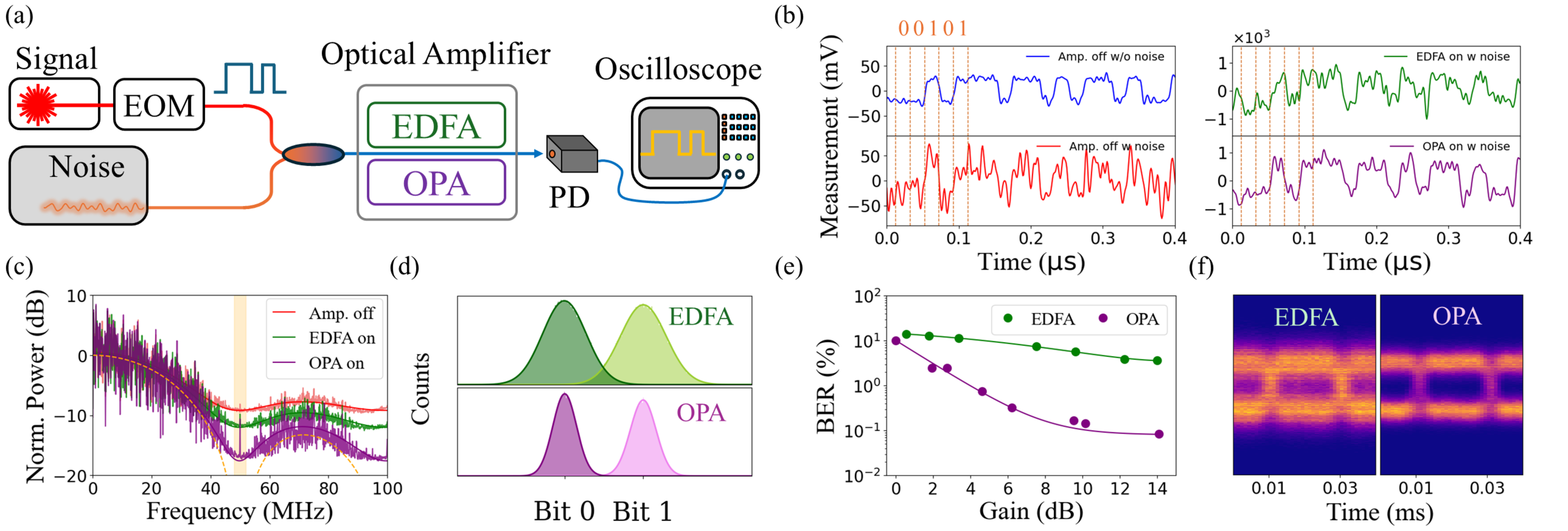}
\caption{
\textbf{Improvement of optical communications with the integrated OPA}
\textbf{a}, Experimental schematic. An on-off keying signal is generated using an intensity electro-optic modulator (EOM). After mixing with broadband noise generated from unseeded EDFA, optical signals are amplified by either EDFA or OPA. The optical signal is detected with a fast photodiode and an oscilloscope. 
\textbf{b}, Time-domain measurement of the optical signal before mixing noise (blue), after mixing noise and before amplification (red), after mixing noise and EDFA amplification (green), and after mixing noise and OPA amplification (purple).
\textbf{c}, Normalized spectrum of photodiode signals without amplification (red), with EDFA amplification (green), and OPA amplification (purple). The gains of EDFA and OPA are both 14~dB. The spectrum is calculated from time-domain signal through fast Fourier transformation. The spectrum is normalized to the power at zero frequency. The shaded area indicates the regime the power is dominated by noise, as the modulation spectrum vanishes.
\textbf{d} Statistical distributions of photodetector output for bit 0 and bit 1 with EDFA and OPA amplifications respectively. Distributions are fitted with Gaussian distribution (see methods). The overlap between bit-0 and bit-1 distributions is proportional to the bit error rate.
\textbf{e} Bit error rate with different EDFA and OPA gain values.
\textbf{f} Eye diagrams of received optical signals after EDFA and OPA amplification with 14~dB gain.
}
\label{fig4}
\end{figure*}

Next, we encode pseudo binary data into the input optical signal with an intensity electro-optic modulator based on square-wave on-off keying (Fig.~\ref{fig4}a). The input optical SNR without modulation is kept at 12~dB, and the modulation frequency is $f_\mathrm{m}$=50~MHz.
We can clearly observe the improvement of the data quality in the time domain with the OPA amplification, but not the EDFA amplification (Fig.~\ref{fig4}b). This is further quantified in the frequency domain through the Fourier transformation of the time-domain data (Fig.~\ref{fig4}c). The total electronic spectrum consists of the sinc-squared function from the ideal square-wave modulation spectrum $S(f) = \mathrm{sinc}^2(\pi f/f_\mathrm{m})$ and the constant background from the input optical noise (Supplementary Section~2). At frequency $f=f_\mathrm{m}$, the modulation spectrum vanishes $S(f_\mathrm{m})=0$ and the total spectrum is dominated by the noise background   (shaded area in Fig.~\ref{fig4}c).
With a similar signal gain around 14~dB, the noise background with the OPA amplification is 5.6~dB lower than EDFA amplification (Fig.~\ref{fig4}c).

The advantage of noise suppression can be alternatively verified by bit error rate (BER) (See Methods). The statistical distributions of the demodulated bit 0 and bit 1 are shown in Fig.~\ref{fig4}d. The overlap between bit-0 and bit-1 distributions is significantly larger with EDFA than OPA.
With EDFA, BER has a marginal improvement from 0.1\% to 0.04\% at 14~dB gain (Fig.~\ref{fig4}e). With 14~dB OPA gain, BER is improved dramatically by over two orders of magnitude from 0.1\% to 0.0008\%. This is also supported by the wider eye diagram with OPA amplification, which shows a clear distinction between bit-0 and bit-1 rails (Fig.~\ref{fig4}f)




In summary, we have demonstrated a highly efficient integrated OPA based on single-pass TFLN waveguides. By employing the adapted poling technique to mitigate nanoscale inhomogeneities, we achieve an on-chip gain of 23.5 dB with a pump power as low as 110~mW. Net gain over 10~dB is demonstrated with 3-dB bandwidth up to 120~nm. With optimized edge‑coupler designs achieving insertion losses below 1~dB~\cite{chen2024high}, the net gain is expected to improve further, exceeding 20~dB.
We further verified that the integrated OPA operates in the quantum regime by observing field fluctuations below the classical shot‑noise limit. This capability is essential for deploying integrated OPAs in quantum technologies, including quantum metrology and fault‑tolerant photonic quantum computing~\cite{Menicucci2014,Takeda2019}.
We further demonstrate the practical potential of this high‑efficiency integrated OPA waveguide for improving optical signal‑to‑noise ratios and lowering bit‑error rates in noisy optical communication systems. Combined with their broad amplification bandwidth, these integrated OPAs can significantly enhance the capacity of optical communication links

Our integrated OPA demonstrates state‑of‑the‑art performance across all key optical‑amplifier metrics, including high net gain, excellent power efficiency, broad bandwidth, quantum‑regime noise characteristics, and a compact footprint. Notably, this level of performance is achieved using the simplest possible device architecture—single‑pass straight waveguides. By avoiding complex photonic structures and cavity‑based enhancement, the integrated OPA attains exceptional robustness and reliability for practical deployment.





\textbf{Acknowledgments}
This work was supported in part by the Defense Advanced Research Projects Agency (DARPA) INSPIRED program (HR0011-24-3-0416), Office of Naval Research (N00014-25-1-2130), Air Force Office of Scientific Research (FA9550-24-1-0119), U.S. Department of Energy (Field Work Proposal ERKJ432). L.F. acknowledge the support from Coherent/II-VI Foundation and Sloan Fellowship. 

\vspace{4pt}
\textbf{Author contributions}

The experiments were conceived by Y.C.K and L.F. The device was designed and fabricated by J.H and I.B. Measurements were performed by Y.C.K. Data analysis was conducted by Y.C.K, P.K.C, and L.F. All authors participated in the manuscript preparation. L.F supervised the work.

\vspace{4pt}
\textbf{Competing interests}
P.K.C, and L.F. are involved in developing lithium niobate technologies at AptoLight LLC. The remaining authors declare no competing interests.

\textbf{Methods} 



\textbf{Device fabrication}

A X-cut thin-film lithium niobate wafer with a 600-nm device layer is used to fabricate OPA devices. The propagation direction is aligned with the Y-axis of lithium niobate. First, we measure the device layer thickness with broadband optical reflectance along the designed waveguide location. Next, nickel electrodes for poling are patterned using the lift-off process with electron-beam lithography and polymethyl methacrylate (PMMA) resist. Electrical pulses are applied to nickle electrodes to invert the domain direction of lithium niobate. After nickel electrodes are removed by hydrochloric acid, electron-beam lithography is used to define photonic waveguides in the poling regime with ma-N resist. Then argon-based plasma is used to transfer waveguide patterns into the lithium niobate device
layer with 350 nm etching depth. Next, devices are annealed in nitrogen gas to lower propagation losses. In the end, photonic integrated chips are cleaved to expose waveguide facets.

\textbf{Gain bandwidth simulation}

The bandwidth is calculated through the dispersion simulation. We consider a TFLN rib waveguide with top width of 2.2~$\mu$m, total thickness of 600~nm, and etch depth of 350~nm.
The phase-matching condition, thus the gain, is proportional to
\begin{equation}
\begin{aligned}
    G_0 \propto \mathrm{sinc}^2[(\beta_\mathrm{p}-\beta_\mathrm{s} -\beta_\mathrm{i}-2\pi/\Lambda)L]
\label{eq:pm_condition}
\end{aligned}
\end{equation}
where $\beta_\mathrm{p}(\omega_\mathrm{p})$, $\beta_\mathrm{s}(\omega)$, and $\beta_\mathrm{i}(\omega_\mathrm{p}-\omega)$ are the propagation constants of the pump, signal and idler, respectively, and $\Lambda$ is the poling period at the center frequency ($\omega = \omega_\mathrm{p}/2$), and $L = 14$~mm is the waveguide length. The pump angular frequency $\omega_\mathrm{p}$ is fixed, and the signal angular frequency $\omega$ is scanned. The bandwidth is defined when $G_0$ drops by 3~dB due to dispersion.



\textbf{OPA gain measurement}

In the direct power measurement, the on-chip gain can be calculated as
\begin{equation}
\begin{aligned}
    G =\frac{P_\mathrm{on}-P_\mathrm{bg}}{P_\mathrm{off}}
\label{eq:gain_opa}
\end{aligned}
\end{equation}
where $P_\mathbf{on}$ and $P_\mathbf{off}$ are the measured output signal powers with the pump on and off respectively, and $P_\mathrm{bg}$ is the spontaneous parametric down-conversion noise background generated by the pump without signal input.

To obtain the net gain including fiber-chip-fiber coupling and propagation losses, we also measure the transmission of the integrated OPA device at the signal wavelength near 1550~nm. The total fiber-chip fiber loss at the signal wavelength is measured as 13.5~dB.

To estimate the on-chip pump power, we measure the transmission of the integrated OPA device at the pump wavelength near 775~nm. The total fiber-chip-fiber transmission at the pump wavelength is measured as $T_\mathrm{p}=0.036$. Therefore, the on-chip pump power can be estimated as $P_\mathrm{p}=\sqrt{T_\mathrm{p}} \cdot P_\mathrm{off-chip}$.

\textbf{Squeezing level estimation}

Output voltages of the homodyne photodetector are captured continuously using an oscilloscope at a sampling rate of 2~GS/s. The output voltage distribution within 10~$\mu$s time window (20,000 samples) is fitted by a Gaussian function $A \cdot \exp[-V^2/2\sigma^2]$ with $A$ the normalization factor, $V$ the output voltage and $\sigma^2$ the variance. This is done first with the pump light and input signal light blocked, which gives the shot noise variance $\sigma_0 ^2$. With the pump on, the variance oscillates. The minimum and maximum variances are assumed to be squeezing and anti-squeezing. The squeezing level is obtained from the ratio of between squeezing and shot noise variances $\sigma_\mathrm{sqz} ^2/\sigma_{0} ^2$. 



\textbf{Bit-error rate calculation}

The bit-error rate (BER) in the communication demonstration is calculated from the statistical distributions of received power levels for bit 0 and 1. The distributions of received power levels are fitted using a Gaussian distribution function.:
\begin{equation}
\begin{aligned}
    P(V_{i}) =A e^{-(V_{i}-x_{i})/2\sigma_{i}^2}
\end{aligned}
\label{eq:snr_hist}
\end{equation}
where indicator $i\in \{ 0,1\}$ corresponds to bit 0 and bit 1, $A$ is a normalization factor, $V_i$ is the photodetector output voltage of individual measurements, and $x_i$ and $\sigma^2$ are the mean and variance of the photodetector output voltage respectively. The BER is then estimated by calculating the overlapped regimes between the bit-0 and bit-1 distributions. For bit 0, we have $\mathrm{BER_0}=\frac{1}{2}+\frac{1}{2}\mathrm{erf}(\frac{x_0-x_1}{2\sqrt{2}\sigma_0})$. For bit 1, we have $\mathrm{BER_1}=\frac{1}{2}-\frac{1}{2}\mathrm{erf}(\frac{x_1-x_0}{2\sqrt{2}\sigma_1})$. Then the total BER is calculated as $\mathrm{(BER_0+BER_1)}/2$.

\bibliography{Ref}

\end{document}